\documentclass{iopart}
\usepackage[T1]{fontenc}

 \expandafter\let\csname equation*\endcsname\relax
 \expandafter\let\csname endequation*\endcsname\relax

\usepackage{amsmath,amssymb,bm}
\usepackage{graphicx,subfig}
\usepackage{esint}

\def \beq {\begin{equation}}
\def \edq {\end{equation}}
\def \bes {\begin{subequations}}
\def \eds {\end{subequations}}
\def \beqn {\begin{equation*}}
\def \edqn {\end{equation*}}

\def \dag {\dagger}

\begin{document}

\title{Magnetic-field asymmetry of nonlinear thermoelectric and heat transport}

\author{Sun-Yong Hwang$^{1,2}$, David S\'anchez$^{2,3}$, Minchul Lee$^{4}$ and Rosa L\'opez$^{2,3}$}

\address{$^1$Department of Physics, Pohang University of Science and Technology, Pohang 790-784, Korea\\
$^2$Institut de F\'{\i}sica Interdisciplin\`aria i Sistemes Complexos IFISC (UIB-CSIC),
E-07122 Palma de Mallorca, Spain\\
$ˆ{^3}$Departament de F\'{\i}sica, Universitat de les Illes Balears, E-07122 Palma de Mallorca, Spain\\
$^4$Department of Applied Physics, College of Applied Science, Kyung Hee University, Yongin 446-701, Korea}
\ead{david.sanchez@uib.es}

\begin{abstract}
	Nonlinear transport coefficients do not obey, in general, reciprocity relations.
	We here discuss the magnetic-field asymmetries that arise in thermoelectric and heat transport of mesoscopic systems.
	Based on a scattering theory of weakly nonlinear transport, we analyze the leading-order symmetry parameters in terms of the screening potential response to either voltage or temperature shifts.
	We apply our general results to a quantum Hall antidot system.
	Interestingly, we find that certain symmetry parameters show a dependence on the measurement configuration. 
\end{abstract}

\maketitle

\section{Introduction}

	The nonlinear regime of mesoscopic transport is unique because certain physical effects have no counterparts
at linear response.
	A prominent example is the breakdown of the Onsager-Casimir relations that manifests itself in the differential conductance out of equilibrium~\cite{san04,spi04}.
	The effect is due to asymmetric properties of electron-electron interactions under reversal of an external magnetic field and has been extensively studied in the last decade both theoretically~\cite{san04,spi04,mart06,and06,kal09,her09,lim10,kub11,lud13} and experimentally~\cite{wei06,mar06,let06,zum06,ang07,har08,che09,bra09}.
	These results are relevant to characterize nonlinear rectification phenomena in ballistic conductors~\cite{son98,lin00,sho01,fle02,but03,gon04,hac04}.
	Nonetheless, all these works deal with purely electric transport. Equally interesting is the investigation
of magnetic-field asymmetries of {\em thermoelectric and heat} rectification transport.
	That is the goal we want to accomplish in this work. 

	It is important to distinguish between magnetic-field asymmetries occurring in the linear and the nonlinear regime of transport.
	For two-terminal conductors coupled to equilibrium environments, the linear conductance is always an even function of the magnetic field $B$~\cite{san08}.
	However, under the same conditions the linear thermoelectric coefficient can exhibit $B$ asymmetries if carriers experience inelastic scattering inside the conductor~\cite{sai11,san11,ent12,jac10}.
	As a consequence, the two-terminal thermopower need not be an even function of $B$ and its degree of asymmetry determines the thermodynamic efficiency of a system with broken time-reversal symmetry~\cite{ben11,bra13,bal13,ape13}.
	The asymmetries we discuss here survive in the purely elastic case and appear only in the nonlinear regime of transport.

	A recent experiment by J. Matthews {\it et al.}~\cite{mat12} has detected an asymmetry of the Seebeck coefficients in a multiterminal cross junction when the applied thermal gradient exceeds the linear response limit.
	Intriguingly, the asymmetry depends on the measurement configuration.
	We consider below general expressions for the thermopower magnetoasymmetry and illustrate our method with an explicit calculation of a model system.
	We find that, quite generally, the symmetric and antisymmetric combinations of the nonlinear thermopower are different depending on the specific way that the generated voltage is measured in response to the applied thermal difference.

	Our analysis is based on a scattering theory valid for nonlinear thermoelectric transport~\cite{san13}.
	This approach considers leading-order contributions to the sample screening potential arising not only from an external dc bias~\cite{but93,chr96} but also from applied temperature shifts~\cite{san13}.
	Thus, our self-consistent treatment takes into account charge injectivity~\cite{but93,chr96} and entropic injectivity~\cite{san13} contributions to the charge accumulation that builds up in the conductor out of equilibrium.
	Recently, the scattering approach has been successfully applied to discuss thermodynamic efficiencies and figures of merit beyond linear response~\cite{whi13a,mea13,whi13b}.
	These results are relevant in view of recent works that emphasize nonlinear thermoelectric effects in superlattices~\cite{ven01}, quantum dots~\cite{sta93,sch08,sve13}, molecular junctions~\cite{red10,lei10}, and quantum impurities in the Kondo regime~\cite{boe01,don02,aze12}.

	Furthermore, the theory~\cite{san13} can be extended to account for nonlinear transport of the {\em heat} flow~\cite{lop13}.
	Surprisingly, nonlinear Peltier effects (a heat flow in response to a voltage shift) in phase-coherent conductors have been less explored~\cite{kul94,bog99}. 
	Heat rectification (a nonlinear heat flow in response to a temperature difference~\cite{seg05}) has been investigated in carbon nanotubes~\cite{cha06} and quantum dots~\cite{ruo11}, just to mention a few.
	Therefore, we naturally extend our analysis of magnetic-field asymmetries to the nonlinear heat transport coefficients.
	We show below that the leading-order heat rectification is $B$-asymmetric when the entropic injectivity is not invariant under reversals of the magnetic field.
	Our study thus aims at providing a complete picture of magnetoasymmetries in quantum conductors simultaneously subjected to large electric and thermal gradients.

\section{Theoretical formalism}

	Suppose that a mesoscopic conductor is attached to multiple terminals $\alpha,\beta,\dots$, where each terminal is characterized both by the electrical voltage bias $eV_{\alpha}=\mu_{\alpha}-E_{F}$ ($\mu_{\alpha}$ is the electrochemical potential and $E_{F}$ is the Fermi energy) and by the thermal gradient $\theta_{\alpha}=T_{\alpha}-T$ ($T_{\alpha}$ and $T$ are the reservoir and the background temperature, respectively).
	The electronic and heat transport is completely described by the scattering matrix $s_{\alpha\beta}=s_{\alpha\beta}(E, eU)$, which is in general a function of the carrier energy $E$ and the electrostatic potential $U$ inside the conductor.
	The potential $U=U(\vec{r},\{V_{\gamma}\},\{\theta_{\gamma}\})$ is, in turn, a function of the position $\vec{r}$ and the set of applied voltages $\{V_{\gamma}\}$ and temperature shifts $\{\theta_{\gamma}\}$.
	The charge and heat currents, at lead $\alpha$ from carriers originated from lead $\beta$, are respectively given by $I_{\alpha}=\frac{2e}{h}\sum_{\beta}\int dE A_{\alpha\beta}(E,eU)f_{\beta}(E)$ and ${\cal J}_{\alpha}=\frac{2}{h}\sum_{\beta}\int dE(E-\mu_{\alpha}) A_{\alpha\beta}(E,eU)f_{\beta}(E)$, where $A_{\alpha\beta}=\text{Tr}[\delta_{\alpha\beta}-s_{\alpha\beta}^{\dag}s_{\alpha\beta}]$ and $f_{\beta}(E)=(1+\exp[(E-\mu_{\beta})/k_{B}T_{\beta}])^{-1}$ is the Fermi distribution function in the reservoir $\beta$.
	We focus on the weakly nonlinear regime of transport, for which we expand these currents around the equilibrium state (defined with $\mu_{\alpha}=E_{F}$ and $T_{\alpha}=T$ for all $\alpha$) up to second order in powers of the driving fields $V_{\alpha}$ and $\theta_{\alpha}$:
\begin{align}
I_{\alpha}&=\sum_{\beta}\Big(G_{\alpha\beta}V_{\beta}+L_{\alpha\beta}\theta_{\beta}\Big)
	+\sum_{\beta\gamma}\Big(G_{\alpha\beta\gamma}V_{\beta}V_{\gamma}
	+L_{\alpha\beta\gamma}\theta_{\beta}\theta_{\gamma}
	+2M_{\alpha\beta\gamma}V_{\beta}\theta_{\gamma}\Big),\label{elec}\\
{\cal J}_{\alpha}&=\sum_{\beta}\Big(R_{\alpha\beta}V_{\beta}+K_{\alpha\beta}\theta_{\beta}\Big)
	+\sum_{\beta\gamma}\Big(R_{\alpha\beta\gamma}V_{\beta}V_{\gamma}
	+K_{\alpha\beta\gamma}\theta_{\beta}\theta_{\gamma}+2H_{\alpha\beta\gamma}V_{\beta}\theta_{\gamma}\Big).\label{heat}
\end{align}
	In Refs.~\cite{san13},~\cite{mea13} and~\cite{lop13}, the general expressions for all linear and leading order nonlinear coefficients are derived.
	In order to make this article self-contained, we write out those coefficients in~\ref{appenA}.
	It should be emphasized that the linear response coefficients $G_{\alpha\beta}$, $L_{\alpha\beta}$, $R_{\alpha\beta}$, and $K_{\alpha\beta}$ are evaluated at equilibrium and consequently are independent of the screening potential $U$, while the weakly nonlinear coefficients $G_{\alpha\beta\gamma}$, $L_{\alpha\beta\gamma}$, $M_{\alpha\beta\gamma}$, $R_{\alpha\beta\gamma}$, $K_{\alpha\beta\gamma}$, and $H_{\alpha\beta\gamma}$ do depend on $U$ in response to the applied electrical and thermal biases.

	In a situation not very far from equilibrium, an expansion of $U$ up to the first order suffices to take account of the interactions:
\begin{equation}\label{eq:U}
U=U_{\text{eq}}+\sum_{\alpha}u_{\alpha}V_{\alpha}+\sum_{\alpha}z_{\alpha}\theta_{\alpha},
\end{equation}
where $u_{\alpha}=(\partial U/\partial V_{\alpha})_{\text{eq}}$ and $z_{\alpha}=(\partial U/\partial\theta_{\alpha})_{\text{eq}}$ are the characteristic potentials (CPs) that relate the variation of the internal potential $U$ to voltage and temperature shifts at terminal $\alpha$.
	In equilibrium case where $U=U_{\text{eq}}$, the screening potential $U$ is symmetric with respect to the reversal of an applied magnetic field $B$ due to the fundamental microscopic reversibility principle, i.e., $U_{\text{eq}}(B)=U_{\text{eq}}(-B)$.
	Corresponding magnetic-field \emph{symmetry} of \emph{linear} thermoelectric and heat transport has been shown in Ref.~\cite{but90} based on the scattering approach. 
	However, when the system is driven into the out-of-equilibrium regime, there is no fundamental reason for this magnetic-field symmetry to hold.
	Indeed, the magnetic-field \emph{asymmetry} emerges because the CPs in Eq.~\eqref{eq:U} are in general magnetic-field asymmetric, i.e., $u_{\alpha}(B)\ne u_{\alpha}(-B)$ and $z_{\alpha}(B)\ne z_{\alpha}(-B)$.
	Thus far~\cite{san04}, the nonlinear electrical conductance $G_{\alpha\beta\gamma}$ in the isothermal case has shown the magnetic-field asymmetry since $u_{\alpha}$ (CP describing the voltage response of $U$) is not an even function of the magnetic field.
	We show here that a magnetic-field asymmetry also arises in the isoelectric case in response to pure thermal gradients due to the asymmetric properties of $z_{\alpha}$ (CP describing the thermal response of $U$).

	The electrostatic potential $U$ is self-consistently determined by considering the net charge of the system $q=q_{\text{bare}}+q_{\text{scr}}$.
	The bare charge $q_{\text{bare}}^{\alpha}$ injected from lead $\alpha$ is due both to a voltage imbalance and to a temperature shift in lead $\alpha$; each contribution is respectively described by the particle injectivity~\cite{but93,chr96} $\nu_{\alpha}^{p}(E)=(2\pi i)^{-1}\sum_{\beta}\text{Tr}\big[s_{\beta\alpha}^{\dag}\frac{ds_{\beta\alpha}}{dE}\big]$ and the entropic injectivity~\cite{san13} $\nu_{\alpha}^{e}(E)=(2\pi i)^{-1}\sum_{\beta}\text{Tr}\big[\frac{E-E_{F}}{T}s_{\beta\alpha}^{\dag}\frac{ds_{\beta\alpha}}{dE}\big]$ summing up to give $q_{\text{bare}}=e\sum_{\alpha}(D_{\alpha}^{p}eV_{\alpha}+D_{\alpha}^{e}\theta_{\alpha})$,
with $D_{\alpha}^{p,e}=-\int dE\nu_{\alpha}^{p,e}(E)\partial_{E}f$.
	The screening charge $q_{\text{scr}}$ builds up inside the conductor due to interaction with the injected charges, which we obtain from the response of the internal potential, $\Delta U=U-U_{\text{eq}}$, away from the equilibrium state $U_{\text{eq}}$.
	The random phase approximation implies $q_{\text{scr}}=e^{2}\Pi\Delta U$ where $\Pi$ is the Lindhard function which in the long wavelength limit becomes $\Pi=\int dED(E)\partial_{E}f$, with $D=D(E_{F})$ the sample density of states.
	Then, the net charge response of the system reads
\begin{equation}\label{q}
q=e\sum_{\alpha}(D_{\alpha}^{p}eV_{\alpha}+D_{\alpha}^{e}\theta_{\alpha})+e^{2}\Pi\Delta U,
\end{equation}
and the set of equations for the CPs is closed when we relate this out-of-equilibrium net charge with $\Delta U$ via the Poisson equation, $\nabla^{2}\Delta U=-4\pi q$.
	Importantly, the self-consistent procedure discussed here is also applicable to inhomogeneous fields, i.e., when the potential $U$ is position-dependent, as will be shown below when we apply our general model to a specific system.

	In order to quantify the aforementioned magnetic-field asymmetry in the nonlinear transport regime, we define the symmetry($\Sigma$) and the asymmetry($A$) parameters for $G$, $L$, $R$, and $K$ coefficients appearing in Eqs.~\eqref{elec} and~\eqref{heat}:
\begin{equation}\label{SigmaA}
\Sigma_{\alpha\beta,\gamma\delta}^{X}\equiv
	\frac{X_{\alpha\beta}(B)X_{\gamma\delta}(-B)}{X_{\alpha\beta}^{\text{linear}}(B)X_{\gamma\delta}^{\text{linear}}(-B)},\qquad
A_{\alpha\beta,\gamma\delta}^{X}\equiv\frac{X_{\alpha\beta}(B)}{X_{\gamma\delta}(-B)},
\end{equation}
where $X_{\alpha\beta}$ refers to the differential transport coefficients ${\cal{G}}_{\alpha\beta}$(electric), ${\cal{L}}_{\alpha\beta}$(thermoelectric), ${\cal{R}}_{\alpha\beta}$(electrothermal), and ${\cal{K}}_{\alpha\beta}$(thermal) defined by
\begin{align}
&{\cal{G}}_{\alpha\beta}\equiv~\frac{\partial I_{\alpha}}{\partial V_{\beta}}\bigg|_{\{\theta\}=0}
	=G_{\alpha\beta}+2G_{\alpha\beta\beta}V_{\beta}+\sum_{\epsilon\ne\beta}(G_{\alpha\beta\epsilon}
		+G_{\alpha\epsilon\beta})V_{\epsilon},\\
&{\cal{L}}_{\alpha\beta}\equiv~\frac{\partial I_{\alpha}}{\partial\theta_{\beta}}\bigg|_{\{V\}=0}
	=L_{\alpha\beta}+2L_{\alpha\beta\beta}\theta_{\beta}+\sum_{\epsilon\ne\beta}(L_{\alpha\beta\epsilon}
		+L_{\alpha\epsilon\beta})\theta_{\epsilon},\label{L}\\
&{\cal{R}}_{\alpha\beta}\equiv\frac{\partial{\cal{J}}_{\alpha}}{\partial V_{\beta}}\bigg|_{\{\theta\}=0}
	=R_{\alpha\beta}+2R_{\alpha\beta\beta}V_{\beta}+\sum_{\epsilon\ne\beta}(R_{\alpha\beta\epsilon}
		+R_{\alpha\epsilon\beta})V_{\epsilon},\\
&{\cal{K}}_{\alpha\beta}\equiv\frac{\partial{\cal{J}}_{\alpha}}{\partial\theta_{\beta}}\bigg|_{\{V\}=0}
	=K_{\alpha\beta}+2K_{\alpha\beta\beta}\theta_{\beta}+\sum_{\epsilon\ne\beta}(K_{\alpha\beta\epsilon}
		+K_{\alpha\epsilon\beta})\theta_{\epsilon},
\end{align}
and $X_{\alpha\beta}^{\text{linear}}$ indicates the corresponding linear terms $G_{\alpha\beta}$, $L_{\alpha\beta}$, $R_{\alpha\beta}$, and $K_{\alpha\beta}$.
	Since we consider either an isothermal, i.e., $\{\theta\}=0$, or an isoelectric case, i.e., $\{V\}=0$, the terms $M_{\alpha\beta\gamma}$ and $H_{\alpha\beta\gamma}$ in Eqs.~\eqref{elec} and~\eqref{heat} do not enter into the above definitions.
	Note here that $X_{\alpha\beta}$ contains both linear and nonlinear contributions and in the linear response regime it satisfies $\Sigma_{\alpha\beta,\beta\alpha}^{X}=A_{\alpha\beta,\beta\alpha}^{X}=1$, due to the microscopic reversibility condition $X_{\alpha\beta}^{\text{linear}}(B)=X_{\beta\alpha}^{\text{linear}}(-B)$.
	Thus, a deviation from 1 of these symmetry and asymmetry parameters is indeed an indication of the magnetic-field symmetry breaking in the nonlinear regime.
	In a recent experiment by J. Matthews \emph{et al.}~\cite{mat12}, the authors tested the magnetic-field asymmetry for the thermoelectric coefficient, i.e., ${\cal{L}}_{\alpha\beta}$ [Eq.~\eqref{L}], for which they defined a parameter quite analogous to $\Sigma_{\alpha\beta,\gamma\delta}^{\cal{L}}$ used here to analyze the measured data, except they averaged the coefficient over the magnetic fields.
	It was shown that sufficiently strong thermal gradients may lead to magnetic-field asymmetries and that these asymmetries qualitatively differ between the diagonal ($\Sigma_{\alpha\alpha,\alpha\alpha}^{\cal{L}}$) and the off-diagonal [$\Sigma_{\alpha\beta,\beta\alpha}^{\cal{L}}$ ($\alpha\ne\beta$)] elements.
	
	In addition to $\Sigma_{\alpha\beta,\gamma\delta}^{\cal{L}}$, we also consider the symmetry parameters $\Sigma_{\alpha\beta,\gamma\delta}^{\cal{G}}$, $\Sigma_{\alpha\beta,\gamma\delta}^{\cal{R}}$, and $\Sigma_{\alpha\beta,\gamma\delta}^{\cal{K}}$, which provide analysis tools for measurements of the electrical or the heat currents.
	In parallel with the symmetry parameters, we also define the asymmetry counterparts, $A_{\alpha\beta,\gamma\delta}^{\cal{G}}$, $A_{\alpha\beta,\gamma\delta}^{\cal{L}}$, $A_{\alpha\beta,\gamma\delta}^{\cal{R}}$, and $A_{\alpha\beta,\gamma\delta}^{\cal{K}}$, for completeness.
	The advantage of using the asymmetry parameters is that they provide pure measures of the magnetic-field asymmetry once they deviate from 1.
	For example, in two-terminal case with $V_{1}=V$ and $V_{2}=0$, we find $\Sigma_{11,11}^{\cal{G}}=1+2(G_{111}^{(B)}+G_{111}^{(-B)})V/G_{11}$ and $A_{11,11}^{\cal{G}}=1+2(G_{111}^{(B)}-G_{111}^{(-B)})V/G_{11}$ up to the leading order in $V$ (see~\ref{appenB}).
	Thus, the nonunity of $A_{11,11}^{\cal{G}}\ne1$ is purely due to the magnetic-field asymmetry $G_{111}^{(B)}\ne G_{111}^{(-B)}$ whereas $\Sigma_{11,11}^{\cal{G}}\ne1$ does not guarantee the field asymmetry but indicates the importance of nonlinear effects, a part of which is the magnetic-field asymmetry.
	As shown in this example, to leading order in the external fields, the symmetry parameter $\Sigma$ consists of the symmetric (even) combination between the nonlinear coefficients [$G_{111}^{(B)}$ and $G_{111}^{(-B)}$ in this case] while the asymmetry parameter $A$ is comprised of the asymmetric (odd) combination, explaining the terminologies.
	If we define the symmetry parameter $\sigma_{\alpha\beta,\gamma\delta}^{X}\equiv X_{\alpha\beta}(B)+X_{\gamma\delta}(-B)$ and the asymmetry parameter $a_{\alpha\beta,\gamma\delta}^{X}\equiv X_{\alpha\beta}(B)-X_{\gamma\delta}(-B)$~\cite{san04}, these are simply related to $\Sigma_{\alpha\beta,\beta\alpha}^{X}$ and $A_{\alpha\beta,\beta\alpha}^{X}$ by $\sigma_{\alpha\beta,\beta\alpha}^{X}/X_{\alpha\beta}^{\text{linear}}=\Sigma_{\alpha\beta,\beta\alpha}^{X}+1$ and $a_{\alpha\beta,\beta\alpha}^{X}/X_{\alpha\beta}^{\text{linear}}=A_{\alpha\beta,\beta\alpha}^{X}-1$ to leading order in $\{V\}$ and $\{\theta\}$.
	But we emphasize that the parameters $\Sigma$ and $A$ which we use here are dimensionless quantities and have direct relevance to the experiments~\cite{mat12}.
	Moreover, these parameters are related to the efficiency of the thermoelectric power generation or the refrigeration~\cite{sai11,ben11,bra13}.
	Thus, the gate-tunability of these parameters, which we demonstrate below for a quantum Hall conductor, can pave the way for controlling the functionality of thermoelectric devices.

\section{Quantum Hall bar}\label{sec:QHB}

\begin{figure}[htbp]
\centering
\includegraphics[width=0.7\textwidth, clip]{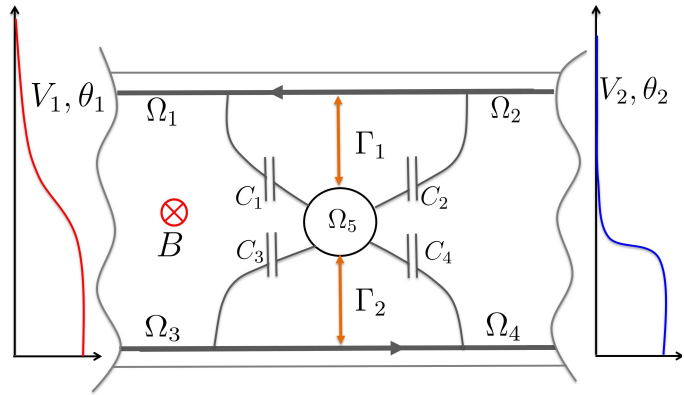}
\caption{Sketch of a quantum Hall bar attached to two reservoirs (1 and 2) with applied voltages $V_{1}$, $V_{2}$ and temperature shifts $\theta_{1}$, $\theta_{2}$.
An antidot ($\Omega_{5}$) is coupled to the quantum Hall edge states with the hybridization widths $\Gamma_{1}$ and $\Gamma_{2}$ and capacitances $C_{1}$, $C_{2}$, $C_{3}$, $C_{4}$.
The antidot level position can be tuned with a top gate potential (not shown here).}\label{fig:sys}
\end{figure}

	Armed with the general formalism described so far, we are now ready to apply it to a specific system; a conductor in the quantum Hall regime coupled to two terminals, as depicted in figure~\ref{fig:sys}.
	We fix the external magnetic field $B$ such that only the lowest Landau level is occupied
(filling factor 1). Hereafter, the magnetic field strength is constant and we only consider the reversal of its direction denoted by $B$ and $-B$.
	An antidot is formed inside the quantum Hall bar by producing a potential hill with a gate control~\cite{for94,kir94}, which can connect two counter-propagating edge states.
	We regard the antidot as a quantum impurity with a Breit-Wigner resonance at $\varepsilon_{0}+eU_{d}(B)$, where $U_{d}(B)$ is the interaction-driven potential shift at the antidot in the presence of magnetic field $B$.
	The upper and the lower edge states are tunnel-coupled to the antidot via hybridization widths $\Gamma_{1}$ and $\Gamma_{2}$, respectively.
	Suppose that the direction of the magnetic field is reversed.
	It follows that the direction of charge flows through the edge states is also reversed due to the chiral nature of the quantum Hall system, and the resonant level at the antidot in this case forms at $\varepsilon_{0}+eU_{d}(-B)$.
	It should be noted that the potential shift $U_{d}$ is in general magnetic-field asymmetric, i.e., $U_{d}(B)\ne U_{d}(-B)$, once the screening effects are incorporated beyond the linear response regime~\cite{san04}.
	This system serves a good test bed for the magnetic-field asymmetry as the symmetry can be broken either through the scattering asymmetry, $\Gamma_{1}\ne\Gamma_{2}$, or through the electrical asymmetry provided the charges on the upper edge interact more strongly with the antidot than those on the lower edge.

	As shown in figure~\ref{fig:sys}, we discretize the conductor potential into five regions $\Omega_{i}$ with $i=1,\dots,5$, where $\Omega_{5}\equiv\Omega_{d}$ denotes the antidot region.
	The potential $U_{i}$ in each region is assumed to be constant and the Coulomb interaction between charges in different regions is described by a capacitance matrix $C_{ij}$~\cite{but93}, making the analytic calculations tractable.
	Despite the simplification, such a discrete local potential model captures the essential physics~\cite{san04,but93,chr96}.
	The region-specific CPs are then given by $u_{i\alpha}=(\partial U_{i}/\partial V_{\alpha})_{\text{eq}}$ and $z_{i\alpha}=(\partial U_{i}/\partial\theta_{\alpha})_{\text{eq}}$, and the net charge response in Eq.~\eqref{q} for each region is related to the capacitance matrix via
\begin{equation}\label{Poisson}
q_{i}=e\sum_{\alpha}(D_{i\alpha}^{p}eV_{\alpha}+D_{i\alpha}^{e}\theta_{\alpha})+e^{2}\Pi_{i}\Delta U_{i}
		=\sum_{j}C_{ij}\Delta U_{j},
\end{equation}
which is a discrete version of the effective Poisson equation.
	The matrix elements $C_{ij}$ are determined by considering the net charge in each region $i$; for instance, we have $C_{11}=-C_{15}\equiv C_{1}$ since $q_{1}=C_{1}(\Delta U_{1}-\Delta U_{5})$, and so on.
	One can determine the potentials $U_{i}$ as a function of the applied voltages and the thermal gradients to obtain the corresponding CPs according to Eq.~\eqref{eq:U}.
	For definiteness, we assume that the density of states for all regions are equal ($D_{i}=D$) and the injectivities in two terminals are symmetric, which amount to $D_{i\alpha}^{p,e}=D^{p,e}$ and $\Pi_{i}=\Pi$.
	We then solve Eq.~\eqref{Poisson} for $\Delta U_{d}=\Delta U_{5}$.
	
	We consider two cases:
(i) the conductor is electrically symmetric, i.e., $C_{i}=C$, but asymmetric in the scattering properties such that $\Gamma_{1}=(1+\eta)\Gamma/2$ and $\Gamma_{2}=(1-\eta)\Gamma/2$, and
(ii) the scattering is symmetric, i.e., $\Gamma_{1}=\Gamma_{2}$, but electrically asymmetric, i.e., $C_{1}=C_{2}=(1+\xi)C$ and $C_{3}=C_{4}=(1-\xi)C$.
	In both cases, the asymmetry is described with a parameter ($\eta$ or $\xi$).
	A little algebra gives $\Delta U_{d}=u_{1}V_{1}+u_{2}V_{2}+z_{1}\theta_{1}+z_{2}\theta_{2}$ and the corresponding CPs
\begin{align}
&u_{1}(B)=u_{2}(-B)=\left\{
	\begin{array}{ll}
		\frac{1}{2}+\eta c_{\text{sc}}\\
		\frac{1}{2}+\xi c_{\text{el}}
	\end{array}\right.,\quad\quad
u_{1}(-B)=u_{2}(B)=\left\{
	\begin{array}{ll}
		\frac{1}{2}-\eta c_{\text{sc}}\\
		\frac{1}{2}-\xi c_{\text{el}}
	\end{array}\right.,\label{CPu}\\
&z_{1}(B)=z_{2}(-B)=\frac{D^{e}}{eD^{p}}u_{1}(B),\quad\qquad
z_{1}(-B)=z_{2}(B)=\frac{D^{e}}{eD^{p}}u_{1}(-B),\label{CPz}
\end{align}
where the terms $\eta c_{\text{sc}}$ and $\xi c_{\text{el}}$ display the results of the two respective cases, (i) scattering asymmetry and (ii) electrical asymmetry,
\begin{equation*}
c_{\text{sc}}=\bigg(2+\frac{4\pi CD^{p}\Gamma}{r(C-e^{2}\Pi)}\bigg)^{-1},\quad
c_{\text{el}}=\frac{-\pi e^{2}\Pi D^{p}C\Gamma t}{(C-e^{2}\Pi)[2\pi CD^{p}\Gamma+r(C-e^{2}\Pi)]}.
\end{equation*}
	Here $r=1-t=\Gamma_{1}\Gamma_{2}/|\Lambda|^{2}$ is the Breit-Wigner reflection ($t$: transmission) probability through the antidot evaluated at equilibrium, with $\Lambda=E_{F}-\varepsilon_{0}+i\Gamma/2$.
	As shown in Eqs.~\eqref{CPu} and \eqref{CPz}, the two asymmetry factors $\eta$ and $\xi$ play qualitatively the same role in the resultant CPs.

	In Eq.~\eqref{CPu}, we firstly note that the sum rule for $u_{\alpha}$ due to gauge invariance [see Eq.~\eqref{gauge_u} in~\ref{appenC}] is indeed satisfied for each direction of the magnetic field as should be: $u_{1}(B)+u_{2}(B)=u_{1}(-B)+u_{2}(-B)=1$.
	One may also note that $\sum_{\alpha}z_{\alpha}=D^{e}/eD^{p}$ is satisfied in Eq.~\eqref{CPz}, but this result is due only to our assumption of equivalent injectivites ($D_{i\alpha}^{p,e}=D^{p,e}$) and in general there is no reason for such a sum rule for $z_{\alpha}$ to exist.
	Importantly, the CPs are generally magnetic-field asymmetric, i.e., $u_{\alpha}(B)\ne u_{\alpha}(-B)$ and $z_{\alpha}(B)\ne z_{\alpha}(-B)$.
	We argue below that the latter asymmetry for $z_{\alpha}$ can explain the recently reported observation of a temperature driven asymmetry beyond linear response~\cite{mat12}.
	It is also important to point out the property $u_{1}(\pm B)=u_{2}(\mp B)$ and $z_{1}(\pm B)=z_{2}(\mp B)$ in Eqs.~\eqref{CPu} and \eqref{CPz}, which can be attributed to the chiral nature of the quantum Hall system.

	The symmetry($\Sigma$) and the asymmetry($A$) parameters defined in Eq.~\eqref{SigmaA} are readily evaluated with the CPs in Eqs.~\eqref{CPu} and \eqref{CPz}.
	The general expressions of these parameters for a generic two-terminal quantum conductor are written in~\ref{appenB}.
	Equations~\eqref{appenB:G}, \eqref{appenB:L}, \eqref{appenB:R}, and \eqref{appenB:K} show that all of the symmetry and the asymmetry parameters can deviate from 1 indicating the importance of nonlinear interactions to leading order in $V$ and $\theta$ because deviations from 1 clearly depend on the CPs.
	Let us now apply the CPs in Eqs.~\eqref{CPu} and \eqref{CPz} evaluated for our quantum Hall system to the general expressions given by Eqs.~\eqref{appenB:G}, \eqref{appenB:L}, \eqref{appenB:R}, and \eqref{appenB:K}.

	Firstly, we consider the off-diagonal asymmetry parameters $A_{\alpha\beta,\beta\alpha}^{X}$ in Eqs.~\eqref{AsymG12}, \eqref{AsymL12}, \eqref{AsymR12}, and \eqref{AsymK12} as well as the electric symmetry parameter $\Sigma_{11,11}^{\cal{G}}$ in Eq.~\eqref{SymG11}.
	We find
\begin{equation}\label{const}
\Sigma_{11,11}^{\cal{G}}=A_{12,21}^{\cal{G}}=A_{12,21}^{\cal{L}}=A_{12,21}^{\cal{R}}=A_{12,21}^{\cal{K}}=1.
\end{equation}
	This constancy is in principle unexpected and stems from the property $u_{1}(\pm B)=u_{2}(\mp B)$ and $z_{1}(\pm B)=z_{2}(\mp B)$.
	Physically, this originates from the fact that our system considered in Fig.~\ref{fig:sys} with $C_{1}=C_{2}$ and $C_{3}=C_{4}$ remain invariant under the simultaneous transformations $B\to-B$ and $V\to-V$.
	In addition to this chirality, the gauge invariance condition ($\sum_{\alpha}u_{\alpha}=1$) plays a role for the derivation of $\Sigma_{11,11}^{\cal{G}}=1$ because $u_{1}(B)+u_{1}(-B)=u_{1}(B)+u_{2}(B)=1$ holds which applies to Eq.~\eqref{SymG11}.
	One can interpret the result as follows: the imposed chirality in the system cancels out the magnetic-field asymmetry and recovers the reciprocity even if weakly nonlinear screening effects are taken into account.
	
	More interestingly, we find that the response of the symmetry parameters for both thermoelectric($\cal{L}$) and thermal($\cal{K}$) coefficients depend on the lead indices:
\begin{subequations}
\begin{align}
&\Sigma_{11,11}^{\cal{L}}=1+c_{1}^{\cal{L}}(2\theta/T),\qquad
\Sigma_{12,21}^{\cal{L}}=1+c_{2}^{\cal{L}}(2\theta/T),\\
&\Sigma_{11,11}^{\cal{K}}=1+c_{1}^{\cal{K}}(2\theta/T),\qquad
\Sigma_{12,21}^{\cal{K}}=1+c_{2}^{\cal{K}}(2\theta/T),
\end{align}
\end{subequations}
[see Eqs.~\eqref{SymL11}, \eqref{SymL12}, \eqref{SymK11}, and \eqref{SymK12}].
	The different tendencies between $\Sigma_{11,11}^{\cal{L}}$ and $\Sigma_{12,21}^{\cal{L}}$ as a function of the thermal gradient $\theta$ has been experimentally observed in an asymmetric multiterminal junction~\cite{mat12}.
	Remarkably, this is a high-temperature effect since at $k_{B}T\to0$ we find $\Sigma_{11,11}^{\cal{L}}=\Sigma_{12,21}^{\cal{L}}=\Sigma_{11,11}^{\cal{K}}=\Sigma_{12,21}^{\cal{K}}=1+2\theta/T$, independently of the system parameters.

\begin{figure}[hbtp]
\centering
\includegraphics[width=1.1\textwidth, clip]{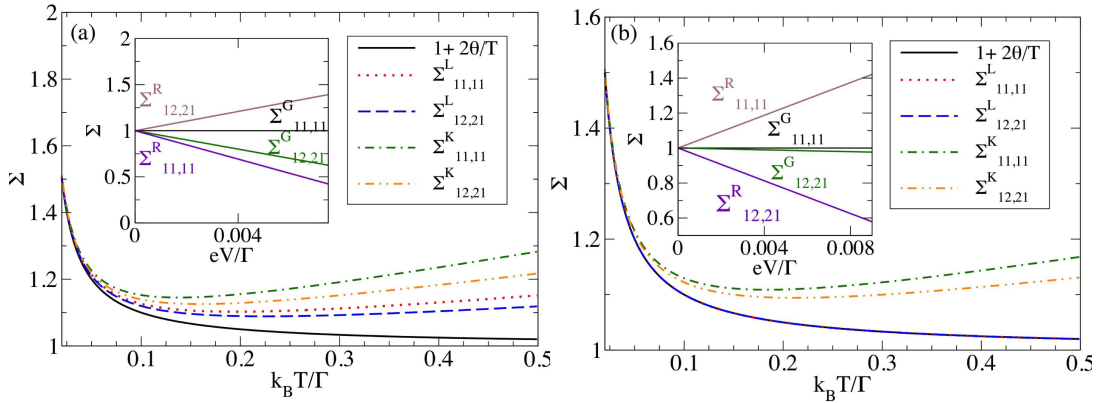}
\caption{Symmetry parameters for the thermoelectric($\Sigma^{\cal{L}}$) and the thermal($\Sigma^{\cal{K}}$) coefficients as a function of the background temperature $T$. The thermal gradient is fixed with $k_B\theta/\Gamma=0.01$.
(a) and (b) show the two distinctive cases of the antidot resonance level $\varepsilon_{0}=0$ and $|\varepsilon_{0}|=\Gamma/2\sqrt3$.
In (b), a merging of the parameters $\Sigma_{11,11}^{\cal{L}}=\Sigma_{12,21}^{\cal{L}}=1+2\theta/T$ is clearly shown. 
 Insets show the voltage dependence of the symmetry parameters $\Sigma^{\cal{R}}$(electrothermal) and $\Sigma^{\cal{G}}$(electric) with $k_{B}T/\Gamma=0.05$.
Left inset corresponds to the case $\varepsilon_0=0.02\Gamma$.
The right inset displays the case $|\varepsilon_{0}|=\Gamma/2\sqrt3$ in which a single constant description for the electrothermal symmetry parameters, i.e., $1-\Sigma_{12,21}^{\cal{R}}=\Sigma_{11,11}^{\cal{R}}-1$, is shown.
We here use $\eta=0.5$ and $E_F=0$ without loss of qualitative generality.
}\label{fig2}
\end{figure}

	We show in Fig.~\ref{fig2} an analysis of the symmetry parameters $\Sigma$ for the various responses.
	In Fig.~\ref{fig2}(a), we first observe a difference between $\Sigma_{11,11}^{\cal{L}}$ and $\Sigma_{12,21}^{\cal{L}}$ at high temperatures.
	Indeed, one can see in Eqs.~\eqref{SymL11} and \eqref{SymL12} that the difference between the symmetry parameters $\Sigma_{11,11}^{\cal{L}}$ and $\Sigma_{12,21}^{\cal{L}}$ for the differential thermoelectric conductance arises from $z_{1}(B)$ and $z_{2}(B)$ incorporated in each parameter, where these CPs characterize the nonlinear thermal responses due to the different leads and in general $z_{1}(B)\ne z_{2}(B)$.
	Our model also predicts that the distinction can also be observed between the thermal symmetry parameters $\Sigma_{11,11}^{\cal{K}}$ and $\Sigma_{12,21}^{\cal{K}}$ [see Fig.~\ref{fig2}(a)] when one measures the heat currents.
	In our quantum Hall system, we find that the diagonal elements $\Sigma_{11,11}^{\cal{L}}$ and $\Sigma_{11,11}^{\cal{K}}$ are totally independent of (i) the scattering asymmetry factor $\eta$ and (ii) the electrical asymmetry factor $\xi$ because $z_{1}(B)+z_{1}(-B)=z_{1}(B)+z_{2}(B)=D^{e}/eD^{p}$ in Eqs.~\eqref{SymL11} and \eqref{SymK11}.
	We digress a little bit and mention that the independence from $\eta$ and $\xi$ is also observed for the diagonal electrothermal element $\Sigma_{11,11}^{\cal{R}}$ in Eq.~\eqref{SymR11}.
	Thus, in our quantum Hall system, eight parameters $\Sigma_{11,11}^{X}$ and $A_{12,21}^{X}$ for all $X={\cal{G}}$, ${\cal{L}}$, ${\cal{R}}$, ${\cal{K}}$, are independent of the scattering asymmetry($\eta$) and the electrical asymmetry($\xi$) factors due to the chiral nature; five of which are manifestly magnetic-field symmetric as already shown in Eq.~\eqref{const}.
	In contrast, the off-diagonal elements $\Sigma_{12,21}^{\cal{L}}$ and $\Sigma_{12,21}^{\cal{K}}$ depend on the asymmetry factors $\eta$ or $\xi$ since the leading order nonlinear terms in Eqs.~\eqref{SymL12} and \eqref{SymK12} include $z_{1}(-B)+z_{2}(B)=2z_{1}(-B)=2z_{2}(B)$.
	When $\eta=\xi=0$, however, the distinction between the diagonal and the off-diagonal elements disappers, i.e., $\Sigma_{11,11}^{\cal{L}}=\Sigma_{12,21}^{\cal{L}}$ and $\Sigma_{11,11}^{\cal{K}}=\Sigma_{12,21}^{\cal{K}}$.
	Therefore, an asymmetry present in the system is crucial to observe this difference.
	This is consistent with the asymmetric scattering used in the experiment~\cite{mat12}.
	We note in passing that, even with nonzero $\eta$ or $\xi$, our analytic results suggest that we can gate-tune the antidot resonance level $\varepsilon_{0}$ to make $c_{1}^{\cal{L}}=c_{2}^{\cal{L}}=1$ (when $|\varepsilon_{0}-E_{F}|=\Gamma/2\sqrt3$) hence recovering the universality of the thermoelectric coefficients, i.e., $\Sigma_{11,11}^{\cal{L}}=\Sigma_{12,21}^{\cal{L}}=1+2\theta/T$.
	This case is precisely shown in Fig.~\ref{fig2}(b) where $|\varepsilon_{0}-E_{F}|=\Gamma/2\sqrt{3}$.  
	However, this is not the case for the heat current counterparts $\Sigma_{11,11}^{\cal{K}}$ and $\Sigma_{12,21}^{\cal{K}}$ and a parameter tuning by means of the antidot top-gate cannot be achieved [see Fig.~\ref{fig2}(b)].

\begin{figure}[hbtp]
\centering
\includegraphics[width=0.9\textwidth, clip]{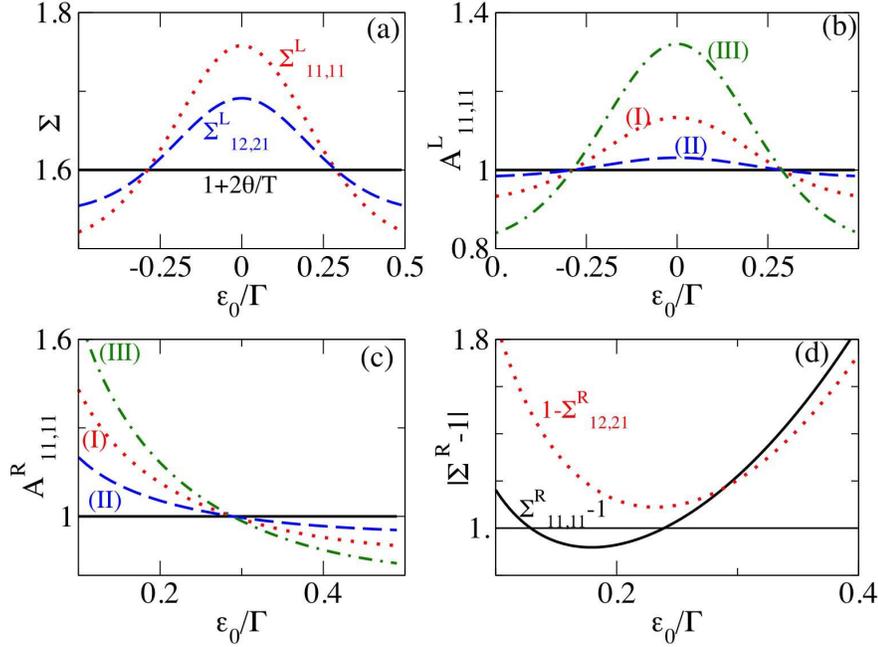}
\caption{Gate tunability of the several parameters.
At $|\varepsilon_{0}|=\Gamma/2\sqrt3\approx0.28\Gamma$, we observe (a) $\Sigma_{11,11}^{\cal{L}}=\Sigma_{12,21}^{\cal{L}}=1+2\theta/T$ (b) $A_{11,11}^{\cal{L}}=1$ (c) $A_{11,11}^{\cal{R}}=1$ and (d) $1-\Sigma_{12,21}^{\cal{R}}=\Sigma_{11,11}^{\cal{R}}-1=c^{\cal{R}}(eV/\Gamma)$.
The fitting parameters are respectively used in (a) $k_{B}T=0.1\Gamma,~k_{B}\theta=0.03\Gamma$, $\eta=0.5$ and (d) $k_{B}T=eV=0.1\Gamma$, $\eta=0.5$.
In (b) [(c)], the three cases refer to
(I) $k_{B}T=0.2\Gamma,~k_{B}\theta=0.03\Gamma~(eV=0.06\Gamma)$ , $\eta=0.5$
(II) $k_{B}T=0.1\Gamma,~k_{B}\theta=0.01\Gamma~(eV=0.02\Gamma)$, $\eta=0.7$
(III) $k_{B}T=0.3\Gamma,~k_{B}\theta=0.04\Gamma~(eV=0.08\Gamma)$, $\eta=0.6$, respectively.
In (c) and (d), we show a small interval around the resonance energy where the thermoelectric and electrothermal asymmetry parameters coincide.
}\label{fig3}
\end{figure}
	There is one more category of parameters whose deviations from the magnetic-field symmetry are directly proportional to either (i) the scattering asymmetry $\eta$ or (ii) the electrical asymmetry $\xi$; these are $\Sigma_{12,21}^{\cal{G}}=1-c^{\cal{G}}(eV/\Gamma)$, $A_{11,11}^{\cal{G}}=1+c^{\cal{G}}(eV/\Gamma)$, $A_{11,11}^{\cal{L}}=1+c_{A}^{\cal{L}}(2\theta/T)$, $A_{11,11}^{\cal{R}}=1+c_{A}^{\cal{R}}(eV/\Gamma)$, and $A_{11,11}^{\cal{K}}=1+c_{A}^{\cal{K}}(2\theta/T)$, in which we find $c^{\cal{G}}=c_{A}^{\cal{L}}=c_{A}^{\cal{R}}=c_{A}^{\cal{K}}=0$ when $\eta=\xi=0$.
	Hence the magneto-asymmetry of these parameters is originated only from the underlying asymmetry in the quantum Hall antidot.
	Note that $\Sigma_{12,21}^{\cal{G}}$ and $A_{11,11}^{\cal{G}}$ are described by a single constant $c^{\cal{G}}$ with opposite signs of the deviation in response to the voltage.
	We again find the gate-tunability such that $A_{11,11}^{\cal{L}}=A_{11,11}^{\cal{R}}=1$ when $|\varepsilon_{0}-E_{F}|=\Gamma/2\sqrt3$ even with nonzero $\eta$ and $\xi$ [see Figs.~\ref{fig3}(b) and \ref{fig3}(c)], which implies that the magnetic-field symmetry can be recovered by adjusting the antidot level.
	Interestingly, this happens at the same resonant level where the universal behavior $\Sigma_{11,11}^{\cal{L}}=\Sigma_{12,21}^{\cal{L}}=1+2\theta/T$ of the thermoelectric symmetry parameters is recovered as explained above.
	
	Finally, we explain the electrothermal symmetry parameters $\Sigma_{11,11}^{\cal{R}}=1+c_{1}^{\cal{R}}(eV/\Gamma)$ and $\Sigma_{12,21}^{\cal{R}}=1-c_{2}^{\cal{R}}(eV/\Gamma)$, describing the voltage response of the magnetic-field asymmetry in heat current measurements, where the latter($\Sigma_{12,21}^{\cal{R}}$) in general depends on $\eta$ or $\xi$ while the former($\Sigma_{11,11}^{\cal{R}}$) does not.
	In the insets of Fig.~\ref{fig2} and Fig.~\ref{fig3}(d), we clearly show that for the particular case where $|\varepsilon_{0}-E_{F}|=\Gamma/2\sqrt3$, we have $c^{\cal{R}}=c_{1}^{\cal{R}}=c_{2}^{\cal{R}}$ yielding $1-\Sigma_{12,21}^{\cal{R}}=\Sigma_{11,11}^{\cal{R}}-1$.
	Besides, at this gate position $\Sigma_{11,11}^{\cal{G}}\approx\Sigma_{12,21}^{\cal{G}}$ as shown in the inset of Fig.~\ref{fig2}(b). 
	
	In Fig.~\ref{fig3}, the aforementioned gate-tunabilities for several parameters are displayed.	
	In our quantum Hall system, we find that the recoveries of the universality for the thermoelectric symmetry parameters $\Sigma_{11,11}^{\cal{L}}=\Sigma_{12,21}^{\cal{L}}=1+2\theta/T$ [Fig.~\ref{fig3}(a)], the magnetic-field symmetry for the diagonal thermoelectric and electrothermal asymmetry parameters $A_{11,11}^{\cal{L}}=A_{11,11}^{\cal{R}}=1$ [Figs.~\ref{fig3}(b) and \ref{fig3}(c)], and the merging into a single constant $1-\Sigma_{12,21}^{\cal{R}}=\Sigma_{11,11}^{\cal{R}}-1=c^{\cal{R}}(eV/\Gamma)$ [Fig.~\ref{fig3}(d)] occur at the same resonance energy, i.e., $|\varepsilon_{0}-E_{F}|=\Gamma/2\sqrt3$.
	It is remarkable that several distinct symmetry and asymmetry parameters can be tuned by a gate-control of the antidot level.
	In Figs.~\ref{fig3}(b) and \ref{fig3}(c), the (diagonal) asymmetry parameters $A_{11,11}^{\cal{L}}$ and $A_{11,11}^{\cal{R}}$ are shown respectively for three different set of values $\{T, \theta~(V), \eta\}$.
	In any case, we have $A_{11,11}^{\cal{L}}=A_{11,11}^{\cal{R}}=1$ at a certain resonance energy, i.e., $|\varepsilon_{0}-E_{F}|=\Gamma/2\sqrt3$.
	In general, our observed gate-tunability is due to the dependence of the CPs on the
antidot level via the reflection and transmission probabilities [see Eqs. (11) and (12) in which $c_{\text{sc}}$ and $c_{\text{el}}$ can be adjusted via $\varepsilon_{0}$].
	We believe that our results are important because the gate-tunability of the magneto-asymmetry is also of practical importance for the evaluation of thermodynamic efficiencies~\cite{ben11,bra13,bal13}.

\section{Conclusion}
	In conclusion, we have investigated the magnetic-field asymmetry of the thermoelectric and the heat transport of mesoscopic systems in the weakly nonlinear regime.
	Based on the scattering approach, we have determined the transport coefficients in terms of the screening potential up to leading order nonlinearity.
	We have defined the symmetry and the antisymmetry parameters which quantify the magnetic-field asymmetry.
	We have applied our general formalism to a two-terminal quantum Hall antidot system and have shown that either voltage or temperature shift leads to the breakdown of Onsager-Casimir symmetry relations beyond the linear response.
	Intriguingly, the underlying chiral nature of our quantum Hall antidot system gives rise to unusual behaviors such as the recovery and gate-tunability of reciprocity even in the weakly nonlinear regime.
	Motivated by this, it will be also interesting to extend our current work to the quantum spin Hall insulator, in which the spin of the carrier and its momentum are correlated giving rise to the helical nature of the system~\cite{QiZhang}, and analyze if there is any peculiar property due to the underlying helicity.

\section*{Acknowledgments}
S.-Y. Hwang and M. Lee were supported by the National Research Foundation (NRF) grant funded by the Korea government (MEST) (Grant No. 2011-0030790).
D. S\'anchez and R. L\'opez were supported by MINECO Grant No. FIS2011-2352.

\appendix

\section{Linear and nonlinear coefficients}\label{appenA}

	The linear coefficients in Eqs.~\eqref{elec} and~\eqref{heat} read
\begin{align}
&G_{\alpha\beta}=\frac{2e^{2}}{h}\int dE A_{\alpha\beta}(E)\Big(-\frac{\partial f(E)}{\partial E}\Big)
		\approx\frac{2e^{2}}{h}A_{\alpha\beta}(E_{F}),\\
&L_{\alpha\beta}=\frac{2e}{hT}\int dE(E-E_{F})A_{\alpha\beta}(E)\Big(-\frac{\partial f(E)}{\partial E}\Big)
		\approx\frac{2e\pi^{2}k_{B}^{2}T}{3h}\frac{\partial A_{\alpha\beta}(E)}{\partial E}\bigg|_{E=E_{F}},\\
&R_{\alpha\beta}=\frac{2e}{h}\int dE(E-E_{F}) A_{\alpha\beta}(E)\Big(-\frac{\partial f(E)}{\partial E}\Big)
		\approx\frac{2e\pi^{2}k_{B}^{2}T^{2}}{3h}\frac{\partial A_{\alpha\beta}(E)}{\partial E}\bigg|_{E=E_{F}},\\
&K_{\alpha\beta}=\frac{2}{h}\int dE\frac{(E-E_{F})^{2}}{T} A_{\alpha\beta}(E)\Big(-\frac{\partial f(E)}{\partial E}\Big)
		\approx\frac{2\pi^{2}k_{B}^{2}T}{3h}A_{\alpha\beta}(E_{F}),
\end{align}
where $f(E)$ is the Fermi distribution function at equilibrium and the Sommerfeld expansion to leading order in $k_{B}T/E_{F}$ at low temperature is taken in all the last approximations.
	The leading order nonlinear coefficients are given by
\begin{align}
G_{\alpha\beta\gamma}&=\frac{-e^{2}}{h}\int dE\Big(\frac{\partial A_{\alpha\beta}}{\partial V_{\gamma}}
		+\frac{\partial A_{\alpha\gamma}}{\partial V_{\beta}}
		+e\delta_{\beta\gamma}\frac{\partial A_{\alpha\beta}}{\partial E}\Big)\frac{\partial f(E)}{\partial E},\\
L_{\alpha\beta\gamma}&=\frac{e}{h}\int dE\frac{E_{F}-E}{T}\Big(\frac{\partial A_{\alpha\beta}}{\partial \theta_{\gamma}}
		+\frac{\partial A_{\alpha\gamma}}{\partial \theta_{\beta}}
		+\delta_{\beta\gamma}\frac{E-E_{F}}{T}\frac{\partial A_{\alpha\beta}}{\partial E}\Big)
			\frac{\partial f(E)}{\partial E},\\
M_{\alpha\beta\gamma}&=\frac{e^{2}}{h}\int dE\Big(\frac{E_{F}-E}{eT}\frac{\partial A_{\alpha\gamma}}{\partial V_{\beta}}
		-\frac{\partial A_{\alpha\beta}}{\partial \theta_{\gamma}}
		-\delta_{\beta\gamma}\frac{E-E_{F}}{T}\frac{\partial A_{\alpha\beta}}{\partial E}\Big)
			\frac{\partial f(E)}{\partial E},\\
R_{\alpha\beta\gamma}&=\frac{e^{2}}{h}\int dE\Bigg\{\delta_{\alpha\gamma}A_{\alpha\beta}
		+\delta_{\alpha\beta}A_{\alpha\beta}-(E-E_{F})\Big(\frac{\partial A_{\alpha\beta}}{\partial eV_{\gamma}}
		+\frac{\partial A_{\alpha\gamma}}{\partial eV_{\beta}}\Big)\\
&\qquad\qquad\qquad\qquad\qquad\qquad\qquad
-\delta_{\beta\gamma}\Big[(E-E_{F})\frac{\partial A_{\alpha\beta}}{\partial E}+A_{\alpha\beta}\Big]\Bigg\}
		\frac{\partial f(E)}{\partial E},\\
K_{\alpha\beta\gamma}&=\frac{-1}{h}\int dE\frac{(E-E_{F})^{2}}{T}\Bigg\{
		\Big(\frac{\partial A_{\alpha\beta}}{\partial \theta_{\gamma}}
			+\frac{\partial A_{\alpha\gamma}}{\partial \theta_{\beta}}\Big)
			+\delta_{\beta\gamma}\Big[\frac{(E-E_{F})}{T}\frac{\partial A_{\alpha\beta}}{\partial E}
			+\frac{A_{\alpha\beta}}{T}\Big]\Bigg\}\frac{\partial f(E)}{\partial E},\\
H_{\alpha\beta\gamma}&=\frac{-e}{h}\int dE(E-E_{F})\Bigg\{
		\Big(\frac{\partial A_{\alpha\gamma}}{\partial \theta_{\beta}}
			+\frac{(E-E_{F})}{T}\frac{\partial A_{\alpha\beta}}{\partial eV_{\gamma}}
			-\delta_{\alpha\gamma}\frac{A_{\alpha\beta}}{T}\Big)\\
&\qquad\qquad\qquad\qquad\qquad\qquad\qquad
			+\delta_{\beta\gamma}\Big[\frac{(E-E_{F})}{T}\frac{\partial A_{\alpha\beta}}{\partial E}
			+\frac{A_{\alpha\beta}}{T}\Big]\Bigg\}\frac{\partial f(E)}{\partial E}.
\end{align}

	For a practical calculation, we use the WKB approximation valid in the long wavelength limit and make the replacement $\delta/\delta U\to-e\partial/\partial E$.
	Then, one can calculate the voltage and the temperature derivatives provided the characteristic potentials are known since
\begin{subequations}
\begin{align}
\frac{\partial A_{\alpha\beta}}{\partial V_{\gamma}}
	&=\frac{\partial U}{\partial V_{\gamma}}\frac{\delta A_{\alpha\beta}}{\delta U}
	\approx-eu_{\gamma}\frac{\partial A_{\alpha\beta}}{\partial E},\\
\frac{\partial A_{\alpha\beta}}{\partial\theta_{\gamma}}
	&=\frac{\partial U}{\partial\theta_{\gamma}}\frac{\delta A_{\alpha\beta}}{\delta U}
	\approx-ez_{\gamma}\frac{\partial A_{\alpha\beta}}{\partial E}.
\end{align}
\end{subequations}
	
	In a two-terminal setup which we consider in Sec.~\ref{sec:QHB}, we have $A_{11}=A_{22}=-A_{12}=-A_{21}=t(E)$ with $t(E)$ the transmission probability.
	Then, one can find to leading order of the Sommerfeld expansion
\begin{subequations}
\begin{align}
G_{111}&=\frac{e^{3}}{h}\frac{\partial t(E)}{\partial E}\bigg|_{E_{F}}(1-2u_{1}),\\
G_{122}&=\frac{e^{3}}{h}\frac{\partial t(E)}{\partial E}\bigg|_{E_{F}}(2u_{2}-1),\\
G_{211}&=\frac{e^{3}}{h}\frac{\partial t(E)}{\partial E}\bigg|_{E_{F}}(2u_{1}-1),
\end{align}
\end{subequations}
\begin{subequations}
\begin{align}
L_{111}&=\frac{e\pi^{2}k_{B}^{2}}{3h}\bigg[\frac{\partial t(E)}{\partial E}
			-2ez_{1}T\frac{\partial^{2}t(E)}{\partial E^{2}}\bigg]_{E_{F}},\\
L_{122}&=-\frac{e\pi^{2}k_{B}^{2}}{3h}\bigg[\frac{\partial t(E)}{\partial E}
			-2ez_{2}T\frac{\partial^{2}t(E)}{\partial E^{2}}\bigg]_{E_{F}},\\
L_{211}&=-\frac{e\pi^{2}k_{B}^{2}}{3h}\bigg[\frac{\partial t(E)}{\partial E}
			-2ez_{1}T\frac{\partial^{2}t(E)}{\partial E^{2}}\bigg]_{E_{F}},
\end{align}
\end{subequations}
\begin{subequations}
\begin{align}
R_{111}&=-\frac{e^{2}}{h}\bigg[t(E)+
		\frac{\pi^{2}(k_{B}T)^{2}}{6}\frac{\partial^{2}t(E)}{\partial E^{2}}(4u_{1}-1)\bigg]_{E_{F}},\\
R_{122}&=-\frac{e^{2}}{h}\bigg[t(E)+
		\frac{\pi^{2}(k_{B}T)^{2}}{6}\frac{\partial^{2}t(E)}{\partial E^{2}}(3-4u_{2})\bigg]_{E_{F}},\\
R_{211}&=-\frac{e^{2}}{h}\bigg[t(E)+
		\frac{\pi^{2}(k_{B}T)^{2}}{6}\frac{\partial^{2}t(E)}{\partial E^{2}}(3-4u_{1})\bigg]_{E_{F}},
\end{align}
\end{subequations}
\begin{subequations}
\begin{align}
K_{111}&=\frac{\pi^{2}k_{B}^{2}}{3h}\bigg[t(E)-2ez_{1}T\frac{\partial t(E)}{\partial E}\bigg]_{E_{F}},\\
K_{122}&=-\frac{\pi^{2}k_{B}^{2}}{3h}\bigg[t(E)-2ez_{2}T\frac{\partial t(E)}{\partial E}\bigg]_{E_{F}},\\
K_{211}&=-\frac{\pi^{2}k_{B}^{2}}{3h}\bigg[t(E)-2ez_{1}T\frac{\partial t(E)}{\partial E}\bigg]_{E_{F}}.
\end{align}
\end{subequations}

\section{Symmetry and asymmetry parameters in two-terminal case}\label{appenB}

	Following the definitions in Eq.~\eqref{SigmaA}, we evaluate all the symmetry($\Sigma$) and the asymmetry($A$) parameters up to the leading order of biases ($V$ and $\theta$) for a generic two-terminal conductor:
\begin{subequations}\label{appenB:G}
\begin{align}
\Sigma_{11,11}^{\cal{G}}&=1+\frac{2\big(G_{111}^{(B)}+G_{111}^{(-B)}\big)}{G_{11}}V
		=1-2\big(u_{1}^{(B)}+u_{1}^{(-B)}-1\big)\frac{\frac{\partial t(E)}{\partial E}}{t(E)}\Bigg|_{E_{F}}eV,
			\label{SymG11}\\
\Sigma_{12,21}^{\cal{G}}&=1+\frac{2\big(G_{122}^{(B)}+G_{211}^{(-B)}\big)}{G_{12}}V
		=1-2\big(u_{1}^{(-B)}+u_{2}^{(B)}-1\big)\frac{\frac{\partial t(E)}{\partial E}}{t(E)}\Bigg|_{E_{F}}eV,
			\label{SymG12}\\
A_{11,11}^{\cal{G}}&=1+\frac{2\big(G_{111}^{(B)}-G_{111}^{(-B)}\big)}{G_{11}}V
		=1-2\big(u_{1}^{(B)}-u_{1}^{(-B)}\big)\frac{\frac{\partial t(E)}{\partial E}}{t(E)}\Bigg|_{E_{F}}eV,
			\label{AsymG11}\\
A_{12,21}^{\cal{G}}&=1+\frac{2\big(G_{122}^{(B)}-G_{211}^{(-B)}\big)}{G_{12}}V
		=1-2\big(u_{2}^{(B)}-u_{1}^{(-B)}\big)\frac{\frac{\partial t(E)}{\partial E}}{t(E)}\Bigg|_{E_{F}}eV,
			\label{AsymG12}
\end{align}
\end{subequations}
\begin{subequations}\label{appenB:L}
\begin{align}
\Sigma_{11,11}^{\cal{L}}&=1+\frac{2\big(L_{111}^{(B)}+L_{111}^{(-B)}\big)}{L_{11}}\theta
	=1+2\Bigg[1-eT\big(z_{1}^{(B)}+z_{1}^{(-B)}\big)
		\frac{\frac{\partial^{2}t(E)}{\partial E^{2}}}{\frac{\partial t(E)}{\partial E}}\Bigg]_{E_{F}}\frac{\theta}{T},
			\label{SymL11}\\
\Sigma_{12,21}^{\cal{L}}&=1+\frac{2\big(L_{122}^{(B)}+L_{211}^{(-B)}\big)}{L_{12}}\theta
	=1+2\Bigg[1-eT\big(z_{1}^{(-B)}+z_{2}^{(B)}\big)
		\frac{\frac{\partial^{2}t(E)}{\partial E^{2}}}{\frac{\partial t(E)}{\partial E}}\Bigg]_{E_{F}}\frac{\theta}{T},
			\label{SymL12}\\
A_{11,11}^{\cal{L}}&=1+\frac{2\big(L_{111}^{(B)}-L_{111}^{(-B)}\big)}{L_{11}}\theta
	=1-2eT\big(z_{1}^{(B)}-z_{1}^{(-B)}\big)
		\frac{\frac{\partial^{2}t(E)}{\partial E^{2}}}{\frac{\partial t(E)}{\partial E}}\Bigg|_{E_{F}}\frac{\theta}{T},
			\label{AsymL11}\\
A_{12,21}^{\cal{L}}&=1+\frac{2\big(L_{122}^{(B)}-L_{211}^{(-B)}\big)}{L_{12}}\theta
	=1+2eT\big(z_{1}^{(-B)}-z_{2}^{(B)}\big)
		\frac{\frac{\partial^{2}t(E)}{\partial E^{2}}}{\frac{\partial t(E)}{\partial E}}\Bigg|_{E_{F}}\frac{\theta}{T},
			\label{AsymL12}
\end{align}
\end{subequations}
\begin{subequations}\label{appenB:R}
\begin{align}
\Sigma_{11,11}^{\cal{R}}&=1+\frac{2\big(R_{111}^{(B)}+R_{111}^{(-B)}\big)}{R_{11}}V
	=1-\Bigg[\frac{6}{\pi^{2}(k_{B}T)^{2}}\frac{t(E)}{\frac{\partial t(E)}{\partial E}}
		+\Big(2\big[u_{1}^{(B)}+u_{1}^{(-B)}\big]-1\Big)
			\frac{\frac{\partial^{2}t(E)}{\partial E^{2}}}{\frac{\partial t(E)}{\partial E}}\Bigg]_{E_{F}}eV,
			\label{SymR11}\\
\Sigma_{12,21}^{\cal{R}}&=1+\frac{2\big(R_{122}^{(B)}+R_{211}^{(-B)}\big)}{R_{12}}V
	=1+\Bigg[\frac{6}{\pi^{2}(k_{B}T)^{2}}\frac{t(E)}{\frac{\partial t(E)}{\partial E}}
		+\Big(3-2\big[u_{1}^{(-B)}+u_{2}^{(B)}\big]\Big)
			\frac{\frac{\partial^{2}t(E)}{\partial E^{2}}}{\frac{\partial t(E)}{\partial E}}\Bigg]_{E_{F}}eV,
			\label{SymR12}\\
A_{11,11}^{\cal{R}}&=1+\frac{2\big(R_{111}^{(B)}-R_{111}^{(-B)}\big)}{R_{11}}V
	=1-2\big(u_{1}^{(B)}-u_{1}^{(-B)}\big)
		\frac{\frac{\partial^{2}t(E)}{\partial E^{2}}}{\frac{\partial t(E)}{\partial E}}\Bigg|_{E_{F}}eV,
			\label{AsymR11}\\
A_{12,21}^{\cal{R}}&=1+\frac{2\big(R_{122}^{(B)}-R_{211}^{(-B)}\big)}{R_{12}}V
	=1+2\big(u_{1}^{(-B)}-u_{2}^{(B)}\big)
		\frac{\frac{\partial^{2}t(E)}{{\partial E^{2}}}}{\frac{\partial t(E)}{\partial E}}\Bigg|_{E_{F}}eV,
			\label{AsymR12}
\end{align}
\end{subequations}
\begin{subequations}\label{appenB:K}
\begin{align}
\Sigma_{11,11}^{\cal{K}}&=1+\frac{2\big(K_{111}^{(B)}+K_{111}^{(-B)}\big)}{K_{11}}\theta
	=1+2\Bigg[1-eT\big(z_{1}^{(B)}+z_{1}^{(-B)}\big)
		\frac{\frac{\partial t(E)}{\partial E}}{t(E)}\Bigg]_{E_{F}}\frac{\theta}{T},\label{SymK11}\\
\Sigma_{12,21}^{\cal{K}}&=1+\frac{2\big(K_{122}^{(B)}+K_{211}^{(-B)}\big)}{K_{12}}\theta
	=1+2\Bigg[1-eT\big(z_{1}^{(-B)}+z_{2}^{(B)}\big)
		\frac{\frac{\partial t(E)}{\partial E}}{t(E)}\Bigg]_{E_{F}}\frac{\theta}{T},\label{SymK12}\\
A_{11,11}^{\cal{K}}&=1+\frac{2\big(K_{111}^{(B)}-K_{111}^{(-B)}\big)}{K_{11}}\theta
	=1-2eT\big(z_{1}^{(B)}-z_{1}^{(-B)}\big)\frac{\frac{\partial t(E)}{\partial E}}{t(E)}\Bigg|_{E_{F}}\frac{\theta}{T},
			\label{AsymK11}\\
A_{12,21}^{\cal{K}}&=1+\frac{2\big(K_{122}^{(B)}-K_{211}^{(-B)}\big)}{K_{12}}\theta
	=1+2eT\big(z_{1}^{(-B)}-z_{2}^{(B)}\big)\frac{\frac{\partial t(E)}{\partial E}}{t(E)}\Bigg|_{E_{F}}\frac{\theta}{T}.
			\label{AsymK12}
\end{align}
\end{subequations}
	Here, the off-diagonal elements $\Sigma_{12,21}^{X}$ and $A_{12,21}^{X}$ are related to both terminals (1 and 2), and hence are evaluated under simultaneous transformations $B\to-B$ and either driving fields $V\to-V$ or $\theta\to-\theta$, while the diagonal elements $\Sigma_{11,11}^{X}$ and $A_{11,11}^{X}$ are evaluated only with $B\to-B$ since $V_{1}=V$, $\theta_{1}=\theta$, and $V_{2}=\theta_{2}=0$ are fixed.
	Thus, in the limit $B\to0$, $A_{11,11}^{X}=A_{12,21}^{X}=1$ ($X=\cal{G}, \cal{L},\cal{R},\cal{K}$).
	Note that even in this $B\to0$ limit, the symmetry parameters $\Sigma_{11,11}^{X}$ and $\Sigma_{12,21}^{X}$ can have deviations from 1, owing to the nonlinear effects irrelevant to $B$-asymmetry as discussed below Eq. (9) in Sec. 2 of the main text.

\section{Relations between symmetry and asymmetry parameters}\label{appenC}
	
	Due to the charge conservation, i.e., unitarity of the scattering matrix $\sum_{\alpha}A_{\alpha\beta}=\sum_{\beta}A_{\alpha\beta}=0$, we have the sum rules for the coefficients:
\begin{align}
&\sum_{\alpha}G_{\alpha\beta}=\sum_{\beta}G_{\alpha\beta}=\sum_{\alpha}G_{\alpha\beta\gamma}=0,\\
&\sum_{\alpha}L_{\alpha\beta}=\sum_{\beta}L_{\alpha\beta}=\sum_{\alpha}L_{\alpha\beta\gamma}=0,\\
&\sum_{\alpha}R_{\alpha\beta}=\sum_{\beta}R_{\alpha\beta}=0,\\
&\sum_{\alpha}K_{\alpha\beta}=\sum_{\beta}K_{\alpha\beta}=\sum_{\alpha}K_{\alpha\beta\gamma}=0,
\end{align}
that are easily verified from the general expressions given in~\ref{appenA}.
	In addition, the physics must be invariant under the common shift of voltages giving rise to the constraint~\cite{chr96} $e\partial_{E}A_{\alpha\beta}+\sum_{\gamma}\partial_{V_{\gamma}}A_{\alpha\beta}=0$.
	This gauge invariance condition gives additional sum rules for $G_{\alpha\beta\gamma}$ and the characteristic potential $u_{\alpha}$:
\begin{align}
&\sum_{\gamma}(G_{\alpha\beta\gamma}+G_{\alpha\gamma\beta})=0,\\
&\sum_{\alpha}u_{\alpha}=1.\label{gauge_u}
\end{align}
In a two-terminal case, these sum rules correspond to $G_{12}=-G_{11}$, $G_{122}=G_{111}=-G_{211}$, $L_{12}=-L_{11}$, $L_{111}=-L_{211}$, $R_{12}=-R_{11}$, and $K_{12}=-K_{11}$, $K_{111}=-K_{211}$, from which one can relate the symmetry and the asymmetry parameters:

\begin{align}
&\frac{\Sigma_{11,11}^{\cal{G}}+A_{11,11}^{\cal{G}}}{2}=1+\frac{2G_{111}^{(B)}}{G_{11}}V,\qquad
	\frac{\Sigma_{12,21}^{\cal{G}}+A_{12,21}^{\cal{G}}}{2}=1-\frac{2G_{111}^{(B)}}{G_{11}}V,\label{sumG}\\
&\frac{\Sigma_{11,11}^{\cal{L}}+A_{11,11}^{\cal{L}}}{2}=1+\frac{2L_{111}^{(B)}}{L_{11}}\theta,\qquad
	\frac{\Sigma_{12,21}^{\cal{L}}+A_{12,21}^{\cal{L}}}{2}=1-\frac{2L_{122}^{(B)}}{L_{11}}\theta,\\
&\frac{\Sigma_{11,11}^{\cal{R}}+A_{11,11}^{\cal{R}}}{2}=1+\frac{2R_{111}^{(B)}}{R_{11}}V,\qquad
	\frac{\Sigma_{12,21}^{\cal{R}}+A_{12,21}^{\cal{R}}}{2}=1-\frac{2R_{122}^{(B)}}{R_{11}}V,\\
&\frac{\Sigma_{11,11}^{\cal{K}}+A_{11,11}^{\cal{K}}}{2}=1+\frac{2K_{111}^{(B)}}{K_{11}}\theta,\qquad
	\frac{\Sigma_{12,21}^{\cal{K}}+A_{12,21}^{\cal{K}}}{2}=1-\frac{2K_{122}^{(B)}}{K_{11}}\theta.
\end{align}
Note that the right hand sides of Eq.~\eqref{sumG} are written in terms only of $G_{111}$ and $G_{11}$ due to the gauge invariance with respect to voltage shifts.

\end{document}